\documentclass[useAMS,usenatbib]{mn2e}
\usepackage{epsfig}

\makeatletter

\newcommand{\Rmnum}[1]{\expandafter\@slowromancap\romannumeral #1@}
\makeatother

\title[The HI dominated Low Surface Brightness Galaxy KKR17]{The HI Dominated Low Surface Brightness Galaxy KKR17}
\author[M. Lam and Others]{M. I. Lam$^{1,2,3}$\thanks{Email:linminyi@nao.cas.cn}, H. Wu$^{1,3}$\thanks{Email:hwu@bao.ac.cn}, M. Yang$^{1,3}$, Z. -M. Zhou$^{1,3}$, W. Du$^{1,3}$,  and Y. -N. Zhu$^{1,3}$ \\
$^{1}$ National Astronomical Observatories, Chinese Academy of Sciences, Beijing, 100012, China \\
$^{2}$University of Chinese Academy of Sciences, Beijing, 100049, China \\
$^{3}$Key Laboratory of Optical Astronomy, National Astronomical Observatories, Chinese Academy of Sciences}

\begin{document}
\date{Received 2014 August; in original form 2014 August}

\pagerange{\pageref{firstpage}--\pageref{lastpage}} \pubyear{2002}

\maketitle

\label{firstpage}

\begin{abstract}
We present new narrow-band (H$\alpha$ and [O\Rmnum{3}]) imagings and optical spectrophotometry of H\Rmnum{2} regions for a gas-rich low surface brightness irregular galaxy, KKR 17. The central surface brightness of the galaxy is $\mu_0(B)$ = 24.15 $\pm$0.03 mag~sec$^{-2}$. The galaxy was detected by \emph{Arecibo Legacy Fast ALFA survey} (ALFALFA), and its mass is dominated by neutral hydrogen (HI) gas. In contrast, both the stellar masses of the bright H\Rmnum{2} and diffuse stellar regions are small. In addition, the fit to the spectral energy distribution to each region shows the stellar populations of H\Rmnum{2} and diffuse regions are different. The bright H\Rmnum{2} region contains a large fraction of O-type stars, revealing the recent strong star formation, whereas the diffuse region is dominated by median age stars, which has a typical age of $\sim$ 600 Myrs. Using the McGaugh's abundance model, we found that the average metallicity of KKR 17 is 12 + (O/H) = 8.0 $\pm$ 0.1. The star formation rate of KKR 17 is 0.21$\pm$0.04 M$_{\odot}$/yr, which is $\sim$1/5 of our Milky Way's. Based on the analysis results to young stellar clusters in H\Rmnum{2} region, it is found that the bright H\Rmnum{2} region showed two sub-components with different velocities and metallicities. This may be caused by the outflow of massive stars or merging events. 
However, the mechanism of triggering star formation in the H\Rmnum{2} region is still uncertain.

\end{abstract}

\begin{keywords}
galaxies: abundances --- galaxies: evolution --- galaxies: individual (KKR 17) --- galaxies: irregular
\end{keywords}

\section{Introduction}
\label{sec:intro}

Low surface brightness galaxies (LSBGs) are thought to be important baryonic contributor to the universe (see \citealt{Impey1997,Bothun1997} for a review). The initial study by \citet{Freeman1970}, based on spiral galaxies, found that the central surface brightness of disk galaxies was concentrated on a very narrow range. Subsequently, \citet{Disney1976} pointed out the surface brightness resulted from \citet{Freeman1970} may be due to the selection effect and predicted the existence of galaxies fainter than sky background. Indeed, many surveys later have discovered a large number of LSBGs (e.g. \citealt{Schombert1988,Schombert1992,Bothun1992,Caldwell1987,Impey1996}). They span a very wide range in morphology (ranging from dwarfs and irregulars to giant disk galaxies), stellar mass and colors (0.3 $<$ B-V $<$ 1.7) (e.g. \citealt{McGaugh1995,Oneil1997}). Considering their metal content, most LSBGs present low metallicity (see, e.g. \citealt{Skillman1989a,Skillman1989b}), although the metallicities of some red LSBGs have been found to be around the solar metallicity \citep{Bergmann2003}.

Star formation plays a crucial role in the evolution of LSBGs. Usually, star formation in galaxies could be divided into four different kinds of mode: (1) instantaneous star formation, typically in interaction/merging event, (2) normal star formation with relatively higher star formation rate (SFR; about 1$\sim$5 M$_{\odot}$/yr), typically in normal spiral galaxies under gravitational wave density, (3) continuous star formation under very low SFR in long time-scale, typically in dwarf galaxies \citep{Schombert2011} and (4) episodic(sporadic) star formation, star formation do not need to be a continuous process under small fluctuation, typically in LSBGs \citep{McGaugh1994}.  LSBGs are found to have relatively low SFRs, probably an order of magnitude lower than HSBGs \citep{Bothun1997,Kim2007}. Previous studies showed that the evolution of LSBGs was much slower than that of their high surface brightness counterparts (HSBGs), and most of their stellar mass was formed by the third star formation mode. However, \citet{Schombert2001} argued that the weak bursts/interaction may still occur in LSBGs in the recent 5 Gyrs. \citet{Kim2007} found that the LSBGs may experience episodic star formation activities along Hubble time due to gas infalling. Recently, \citet{Schombert2013} and \citet{Schombert2014} both found the star formation mechanism in their sample appeared to be the same as those of HSBGs, and the total stellar mass formation in LSBGs were close to a Hubble time, alternatively, it was a normal star formation mode. All of these results reveal that the star formation activities in LSBGs are still controversial.

\begin{table*}
\centering
\begin{minipage}{150mm}
\caption{Observation Log}\label{tab:log}
\small
\begin{tabular}{lccccccc}
  \hline\noalign{\smallskip}
  \hline\noalign{\smallskip}

Band & Telescope & Instrument & $\lambda_{eff}$ & Exposure Time & Date & FWHM & Pixel Size \\
      &           &            &  ($\mu$m) & (sec) & (UT) & ($\arcsec$) & ($\arcsec$ pixel$^{-1}$) \\
\hline\noalign{\smallskip}
FUV & \it{GALEX} & ... & 0.1516 & 1701.05 & 2009 May 28 & 6.0 & 1.500 \\
NUV & \it{GALEX} & ... & 0.2267 & 1701.05 & 2009 May 28 & 6.0 & 1.500 \\
$u$ & SDSS & ... & 0.3543 & 53.9 & 2003 Jun 22 & 1.4 & 0.396 \\
$g$ & SDSS & ... & 0.4770 & 53.9 & 2003 Jun 22 & 1.4 & 0.396 \\
$r$ & SDSS & ... & 0.6231 & 53.9 & 2003 Jun 22 & 1.4 & 0.396 \\
$i$ & SDSS & ... & 0.7625 & 53.9 & 2003 Jun 22 & 1.4 & 0.396 \\
$z$ & SDSS & ... & 0.9134 & 53.9 & 2003 Jun 22 & 1.4 & 0.396 \\
3.4 \micron & WISE & ... & 3.368 & & 2010 & 6.0 & 1.375 \\
21 cm & Arecibo & L-band & 21 cm & ... & ... & $\sim$3.5$\arcmin$ & ... \\

\hline
\\
& & & Our Observations & & & & \\
\hline
\\
V & Xinglong 2.16m & BFOCS & 0.545 & 2 $\times$ 300 & 2012 Jun 16 & 2.2 & 0.45 \\
R & Xinglong 2.16m & BFOCS & 0.700 & 2 $\times$ 300 & 2012 Jun 17 & 2.2 & 0.45 \\
$[\rm{O}\Rmnum{3}]$-4 & Xinglong 2.16m & BFOCS & 0.516 & 2 $\times$ 1800 & 2012 Jun 16 & 2.2 & 0.45 \\
H$\alpha$-5 &  Xinglong 2.16m & BFOCS & 0.676 & 3600 & 2012 Jun 17 & 2.2 & 0.45 \\
Spectrum (G6) & Xinglong 2.16m & BFOCS & 0.33-0.545 & 3600 & 2013 Mar 10 & 3.0 & 0.45 \\
Spectrum (G7) & Xinglong 2.16m & BFOCS & 0.39-0.67 & 3600 & 2013 Jun 15 & 1.8 & 0.45 \\
Spectrum (G8) & Xinglong 2.16m & BFOCS & 0.58-0.82 & 3600 & 2012 Jun 17 & 2.2 & 0.45 \\

\noalign{\smallskip}\hline
\end{tabular}
\end{minipage}
\end{table*}

In order to better understand the effects on evolution of LSBGs, it is essential to study the star formation properties of nearby LSBGs. Therefore, we selected one typical blue LSBG from Arecibo Legacy Fast ALFA Survey (ALFALFA; \citealt{Giovanelli2005a}), KKR 17, to investigate its possible star formation activity. KKR 17 has one bright, compact knot, and one extended, faint diffuse region. This galaxy was firstly discovered by the Second Palomar Sky Survey (POSS-\Rmnum{2}) films \citep{KKR1999}, and its redshift was confirmed to be 0.0276 \citep{Makarov2003}. In addition, its H\Rmnum{1} flux was measured by using the 100-m radio telescope at Effelsberg \citep{Huchtmeier2000}. The Arecibo ALFALFA survey showed its gas mass was quite high. Based on literature available, KKR 17 is still poorly studied and could be an ideal laboratory to test some of the star formation models. 

In this paper, we explored the global properties of KKR 17, using the multi-wavelength observations from ultraviolet (UV) to near-infrared (NIR), and mainly focusing on the star formation and stellar population. The observations and relevant data reduction are presented in \textsection 2. The main results of our analysis are presented in \textsection 3. Discussions and summary are presented in \textsection 4 and \textsection 5, respectively.

\begin{figure*}
\includegraphics[width=130mm]{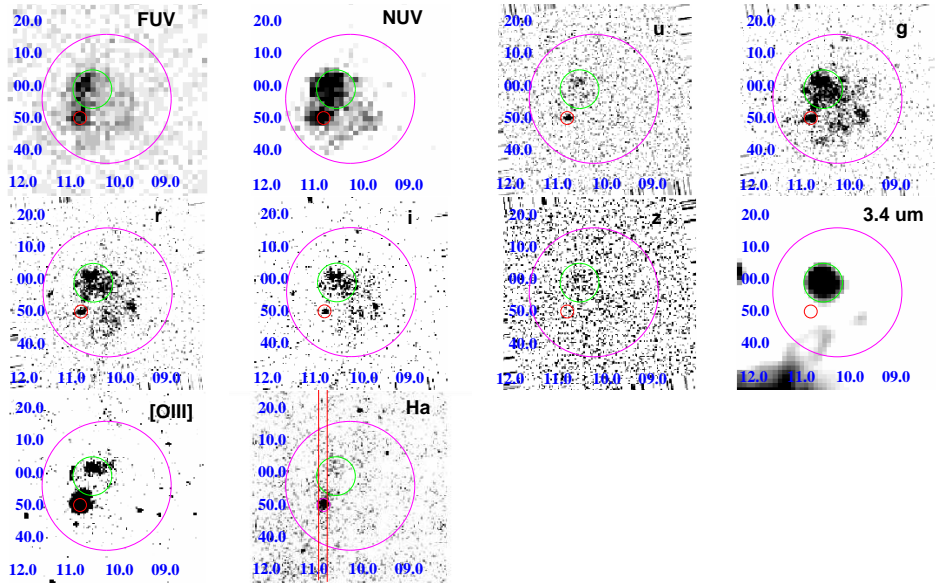}
\caption{The multi-wavelength images of KKR 17 (centered on RA = 15:11:10.2 \& DEC = +11:01:56). From left to right and from top to bottom are FUV, NUV images from GALEX, $ugriz$ images from SDSS, 3.4 \micron~image from WISE and [O\Rmnum{3}], H$\alpha$ narrow-band images from 2.16m telescope at Xinglong observatory. The narrow-band images were already stellar-continuum subtracted. The magenta circles are our aperture circle. The spectra was extracted from region 1 (red circles), the low mass stars are dominated in region 2 (green circles). The red lines in H$\alpha$ image indicate the position of long-slit of our spectrum.
\label{fig:images}}
\end{figure*}

\section{Observation and Data Reduction}
\label{sec:data}
We used the archived far-UV (FUV), Near-UV (NUV) images from \emph{Galaxy Evolution Explorer} (GALEX), optical wide-band images ($ugriz$) from \emph{Sloan Digital Sky Survey} (SDSS), mid-infrared 3.4 \micron~images from \emph{Wide-Field Infrared Survey Explorer} (WISE) and HI data from \emph{Arecibo Legacy Fast ALFA Survey} (ALFALFA). In addition, we have obtained optical narrow-band images and spectra of KKR 17 by using BAO Faint Object Spectrograph and Camera (BFOSC) mounted on 2.16m telescope at Xinglong Observatory, National Astronomical Observatories of Chinese Academy of Science (NAOC). Detailed information is given in Table~\ref{tab:log}.

\subsection{Ultraviolet Images}
The GALEX mission was launched in 2003, and it has surveyed all sky simultaneously in two broad-bands. The effective wavelength of FUV instrument is 1516\AA, and NUV is 2267\AA~\citep{Martin2005}. The field of view (FOV) of GALEX is $\sim$1.2$^{\circ}$ \citep{Morrissey2007}. We obtained the images from GALEX Medium Imaging Survey (MIS; $\sim$1500s), which was designed to cover 1000 $\deg^2$ and reach to $m_{AB} \approx$ 23 mag. This survey could also cover the maximum area of SDSS survey. The keywords of mean sky-background level (SKY) and AB magnitude zero point (ZP) in the header of image were used for sky subtraction and flux calibration, respectively. The final image has a spatial resolution of 6\arcsec~and a pixel size of 1.5\arcsec. 

\subsection{Optical images}
The optical broadband images ($ugriz$) were taken from SDSS data archive server \citep{York2000,Stoughton2002}. 
The background was subtracted from each image before photometry, and the counts were converted to flux densities and magnitudes.

The new images with narrow-band filters (centered at redshifted H$\alpha$ and [O\Rmnum{3}], respectively) and broad-band filters (V- and R-bands) were observed with BFOSC on 2.16m telescope at Xinglong Observatory, NAOC on June 16 2012. The FOV of BFOSC is approximately 8.5\arcmin $\times$9.5\arcmin, with the CCD size of 1130$\times$1230 pixels. One pixel of BFOSC corresponds to $\sim$0.45\arcsec. The continuum-subtracted [O\Rmnum{3}] images included both [O\Rmnum{3}]$\lambda$$\lambda$ 4959 and 5007 in our images. However, The continuum-subtracted `H$\alpha$' images still included a component from [N\Rmnum{2}]$\lambda$$\lambda$6548,6583 emission lines. Fortunately, the doublet [N\Rmnum{2}] lines were very weak in the low metallicity environment, which could contaminate $\sim$5\% fluxes at most to our narrow-band image and is much less the flux calibration error. Therefore, we neglected the [N\Rmnum{2}] emission lines in KKR 17 in our analysis.
 
All standard CCD reductions were performed before astrometric and flux calibrations. The SDSS field stars were used as references for astrometric and flux calibration, and the astrometric calibration accuracy is better than 1\arcsec~in our images. Flux calibrations were converted between SDSS and UBVR$_{\rm{c}}$I$_{\rm{c}}$ magnitude systems, using the conversion coefficients in Lupton(2005)\footnote{http://www.sdss.org/dr7/algorithms/sdssUBVRITransform.html}. The V- and R-band images were used as stellar continuum, which would be subtracted from narrow-band [O\Rmnum{3}] and H$\alpha$ images. We estimated that the error of integrated flux is less than 20\% in the narrow-band images, which was limited by the low surface brightness of the galaxy and the flux calibration error.

\subsection{Infrared Data}
The infrared images of KKR 17 were taken from WISE satellite \citep{Wright2010}. The mission was launched in 2009 and began to survey all sky in 2010. The astrometric accuracy for high signal-to-noise (S/N) images was better than 1.5 \arcsec. The angular resolution was 6.1\arcsec, 6.4\arcsec, 6.5\arcsec and 12.0\arcsec~at the wavelengths of 3.4, 4.6, 12 and 22 \micron, respectively. KKR 17 was only found at 3.4 \micron~due to the contamination of nearby bright stars. Moreover, as shown in Figure~\ref{fig:images}, 3.4 \micron~is also contaminated by the nearby bright stars. Therefore, we need to mask the bright star to perform the aperture photometry in this band. 


\subsection{HI Data}
The \emph{Arecibo Legacy Fast ALFA survey} (ALFALFA; \citealt{Giovanelli2005a}) is the largest blind HI line survey nowadays. The current catalog, `$\alpha$.40', which covered $\sim$40\% of the final targeted sky area \citep{Haynes2011}, contained $\sim$15,000 extragalactic sources in the regions of (1) spring: 07$^h$30$^m$ $<$ RA $<$ 16$^h$30$^m$, +04$^{\circ}$ $<$ DEC $<$ +16$^{\circ}$  and +24$^{\circ}$ $<$ DEC $<$ +28$^{\circ}$ , and (2) fall: 22$^h$ $<$ RA $<$ 03$^h$, +14$^{\circ}$ $<$ DEC $<$ +16$^{\circ}$ and +24$^{\circ}$ $<$ DEC $<$ +32$^{\circ}$. The catalog includes the source information on position, HI fluxes, HI masses, systemic velocities, and HI line width, etc.

The 21 cm line profile of KKR 17 is archived from ALFALFA survey \citep{Giovanelli2005a}. The HI line of KKR 17 is shown as a single-horn profile on the velocity map, which may indicate that it is a face-on or unstable-disk galaxy. The HI mass of KKR 17 is $\sim$4.37$\times$10$^9$ M$_{\odot}$ in 120.8 Mpc \citep{Haynes2011}, which shows the mass of neutral hydrogen is comparable to that in a normal galaxy. The receding velocity of the galaxy is 8283 $\pm$ 4 km/s, which is consistent with the Hubble flow velocity inferred from the redshift (z=0.0277) of our spectral analysis.

\begin{table*}
\centering
\begin{minipage}{150mm}
\caption{Properties of KKR 17}\label{tab:properties}
\end{minipage}
\small
\begin{tabular}{lccc}
  \hline\noalign{\smallskip}
  \hline\noalign{\smallskip}
Parameter & Diffuse region  & H\Rmnum{2} region & Total \\
 \hline\noalign{\smallskip}
Optical RA (J2000) & 15:11:10.5 & 15:11:10.76 & 15:11:10.2 \\
Optical Dec (J2000) & +11:01:59 & +11:01:50.05 & +11:01:56 \\
Radio RA (J2000) & ... & ... & 15:11:07.90 \\
Radio Dec (J2000) & ... & ... & +11:02:37 \\
W50 $^{a}$ & ... & ... & 29 $\pm$ 2 \\
F$_{\rm{H}\Rmnum{1}}$ $^{b}$ & ... & ... & 1.26 $\pm$ 0.05 \\
Distance $^{c}$ & ... & ... & 120.8 $\pm$ 8.5 \\
M$_{\rm{H}\Rmnum{1}}$/M$_{\sun}$ $^{d}$ & ... & ... & 4.37 $\times$ 10$^9$ \\
$\mu_0$(B)  $^{e}$ & ... & ... & 24.15 $\pm$ 0.03  \\
FUV $^{f}$ & 23.29$\pm$0.19 & 21.44$\pm$0.10 & 19.09$\pm$0.06 \\
NUV  & 20.14$\pm$0.08 & 21.64$\pm$0.06 & 19.02$\pm$0.04 \\
$u$ $^{g}$ & 19.54$\pm$0.10 & 20.60$\pm$0.12 & 18.43$\pm$0.15 \\
$g$  & 18.72$\pm$0.02 & 20.22$\pm$0.04 & 17.68$\pm$0.03 \\
$r$  & 18.62$\pm$0.03 & 20.49$\pm$0.07 & 17.49$\pm$0.04 \\
$i$  & 18.48$\pm$0.04 & 20.93$\pm$0.14 & 17.37$\pm$0.06 \\
$z$  & 18.46$\pm$0.15 & 20.88$\pm$0.44 & 17.18$\pm$0.22 \\
$[\rm{O}\Rmnum{3}]$ $^{h}$ & 0.78$\pm$0.08 & 1.26$\pm$0.03 & 1.86$\pm$0.20 \\
H$\alpha$  & 0.33$\pm$0.08 & 0.64$\pm$0.03 & 1.51$\pm$0.26  \\
3.4 $\mu$m  & 19.98$\pm$0.21 & 22.76$\pm$0.59 & 18.78$\pm$0.15 \\

\noalign{\smallskip}\hline
\end{tabular}
 \medskip
\\
(a) The unit of W50: km s$^{-1}$; (b) unit of F$_{\rm{H}\Rmnum{1}}$: Jy km s$^{-1}$; (c) unit of distance: Mpc; (d) calculate the mass with the distance of 120.8 Mpc; (e) unit of $\mu_0$(B): (magsec$^{-2}$; (f) unit of FUV, NUV and 3.4$\mu$m: mag in AB system; (g) unit of $ugriz$: mag in SDSS system; (h) unit of $[\rm{O}\Rmnum{3}]$ and H$\alpha$: 10$^{-14}$ erg/s/cm$^2$. 
\end{table*}

\subsection{Photometry}
Photometries of data in all bands were performed after CCD pre-procedures, which included overscan subtraction, bias subtraction, flat-field correction, cosmic-ray reduction and background subtraction. The IRAF task SURFIT was used for background subtraction. We used the IRAF task PHOT to produce the photometry with the same aperture, and fixed the aperture center at RA = 15:11:10.2, DEC = +11:01:56. The aperture radius was 20\arcsec~for entire analysis. In addition, we selected two different regions to analyze their properties: (1) the `diffuse' region centered on RA = 15:11:10.5, DEC = +11:01:59 with photometric aperture radius of 6\arcsec; (2) the H\Rmnum{2} region centered on RA = 15:11:10.76, DEC = +11:01:50.05, with photometric aperture radius of 2\arcsec. All the photometric results are shown in Table~\ref{tab:properties}.

\begin{figure}
\includegraphics[width=0.99\columnwidth]{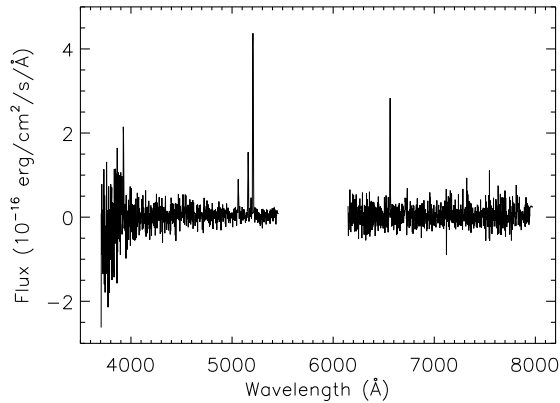}
\caption{The spectra of H\Rmnum{2} region (region 1 in Figure~\ref{fig:images}) of KKR 17. \label{fig:spectra}}
\end{figure}

\subsection{Optical Spectra}
The optical spectra of the H\Rmnum{2} region in KKR 17 were observed on March 10, 2013 (G6+1.8\arcsec), June 15, 2012 (G7+1.8\arcsec) and June 17, 2012 (G8+1.8\arcsec). The observations were also performed by BFOSC on 2.16m telescope. A slit with a width of 1.8\arcsec, combining with three different grisms (G6 from 3600 to 5450\AA, G7 from 4000 to 6700\AA, and G8 from 5800 to 8000\AA), were used for the observation. However, only G6 and G8 spectra were analyzed. The resolution of the grism is $\sim$10\AA. KKR 17 and its H\Rmnum{2} regions cannot be seen by the guide cameras of 2.16 telescope because their flux is quite low. So it is kind of tricky to put the slit properly. We noticed that there was one bright star nearby in the field of view of guide camera, and we estimated the offset between the star and H\Rmnum{2} region through the narrow-band image. The slit was then fixed at the position with the estimated offset relative to the star. The CCD reductions including overscan and bias subtraction, flat-field correction and cosmic-ray reduction were performed before wavelength and flux calibrations. The Fe/Ar lamp was used as the wavelength standard in our spectra, and the Kitt Peak National Observatory IRS standard stars were adopted as flux standard. The optical spectra were shown in Figure~\ref{fig:spectra} and all strong emission lines that we measured were shown in Table~\ref{tab:lines}.

\begin{table}
\centering
\begin{minipage}{80mm}
\caption{Emission-Lines Fluxes and Errors}\label{tab:lines}
\end{minipage}
\small
\begin{tabular}{lcc}
  \hline\noalign{\smallskip}
  \hline\noalign{\smallskip}
Line & Equivalent Width (EW) & Intensity (I{$_\lambda$}) \\
 & ($\AA$) & (10$^{-15}$ erg/s/cm$^2$)\\
\hline
& 2012 Jun 15 & \\
\hline
& upper aperture  & \\
\hline
H${\beta}$ (4861$\AA$) & -249.5 $\pm$ 23 & 1.25 $\pm$ 0.10 \\
$[\rm{O}\Rmnum{3}]$ $\lambda$4959 & -667.5 $\pm$ 32 & 1.82 $\pm$ 0.09 \\
$[\rm{O}\Rmnum{3}]$ $\lambda$5007 & -2429 $\pm$ 645 & 5.13 $\pm$ 0.10 \\
\hline
& bottom aperture  & \\
\hline
H${\beta}$ (4861$\AA$)  &  -552.3 $\pm$ 188& 0.46 $\pm$ 0.10 \\
$[\rm{O}\Rmnum{3}]$ $\lambda$4959  & -427.9 $\pm$ 50 & 0.68 $\pm$ 0.08 \\
$[\rm{O}\Rmnum{3}]$ $\lambda$5007  & -3537 $\pm$ 278& 1.62 $\pm$ 0.09\\
~\\
\hline
\hline
& 2012 Jun 17 & \\
\hline
\\ 
H${\alpha}$ (6563$\AA$) $^{\star}$ & -593.6 $\pm$ 33 & 1.81 $\pm$ 0.13 \\
$[\rm{N}\Rmnum{2}]$ $\lambda$6583 & -27.6 $\pm$ 36 & 0.10 $\pm$ 0.10  
~\\
\hline
\hline
& 2013 Mar 10 & \\
\hline
~\\
$[\rm{O}\Rmnum{2}]$ $\lambda$3727 & -225.9 $\pm$ 11 & 1.59 $\pm$ 0.11 \\
H${\beta}$ (4861$\AA$) $^{\star}$ & -212.2 $\pm$ 72 & 0.80 $\pm$ 0.10 \\
$[\rm{O}\Rmnum{3}]$ $\lambda$4959 & -290.1 $\pm$ 40 & 1.23 $\pm$ 0.10 \\
$[\rm{O}\Rmnum{3}]$ $\lambda$5007 & -610.7 $\pm$ 16 & 3.85 $\pm$ 0.09 \\
~\\

\noalign{\smallskip}\hline
\end{tabular}
 \medskip
\\
$^{\star}$: The high I(H$\beta$)/I(H$\alpha$) ratio of KKR 17 is due to double-components of H$\beta$ emission line. The spectra resolution was limited by the telescope aperture, and it is a small-size telescope in our case. The spectral resolution for 2.16m telescope is only $\sim$10\AA, which is challenging to distinguish the precise sub-structures of KKR 17. Also, the observation seeings on two days were different, which may lead to a small position deviation between two spectra. 
\end{table}

\section{Results}
\label{sec:morpholog & SED}
\subsection{Multiwavelength Morphologies in KKR 17}
KKR 17 is an irregular, gas-rich galaxy. It shows a significant bright H\Rmnum{2} region and a very diffuse, faint region. Figure~\ref{fig:images} showed the different band images from FUV to NIR and the continuum-subtracted [O\Rmnum{3}] and H$\alpha$ images. We separated KKR 17 into two different parts, and discussed their different spectral energy distributions (SEDs) and stellar populations in the following sections. KKR 17 showed different morphologies from UV to NIR, and also centered on different regions in different bands. The H\Rmnum{2} was the brightest region in FUV and NUV, in contrast, the diffuse region was also bright in NUV, but faint in FUV. In addition, in the optical $u$ band which has a low sensitivity, H\Rmnum{2} and diffuse regions showed different features: the diffuse region was barely detectable, but H\Rmnum{2} region was still bright enough for detection. The optical $g$-, $r$-, $i$-band images showed very similar morphologies for KKR 17. This galaxy showed an unstable-disk feature in $g$-, $r$-, and $i$-bands. However, $g$- and $r$-band images are contaminated by the [O\Rmnum{3}] and H$\alpha$ emission, respectively. The narrow-band [O\Rmnum{3}] image with continuum subtracted showed the emission in the bright H\Rmnum{2} knot was relatively stronger than the one in the diffuse region, although the radiation fields in both bright H\Rmnum{2} knot and diffuse region were strong. However, only the narrow-band H$\alpha$ image showed the bright H\Rmnum{2} region, which indicates H\Rmnum{2} knot are undergoing relatively strong star formation activity. The low sensitivity optical $z$-band image was faint in both two regions. The 3.4 \micron~emission is dominated by the ones from late-type stars centered at RA = 15:11:10.5, DEC = +11:01:59, which may be the real galactic center. The disappearance of H\Rmnum{2} region in 3.4 \micron~also revealed that the mass of this region was relatively small.

\begin{figure}
\includegraphics[width=0.9\columnwidth,angle=90]{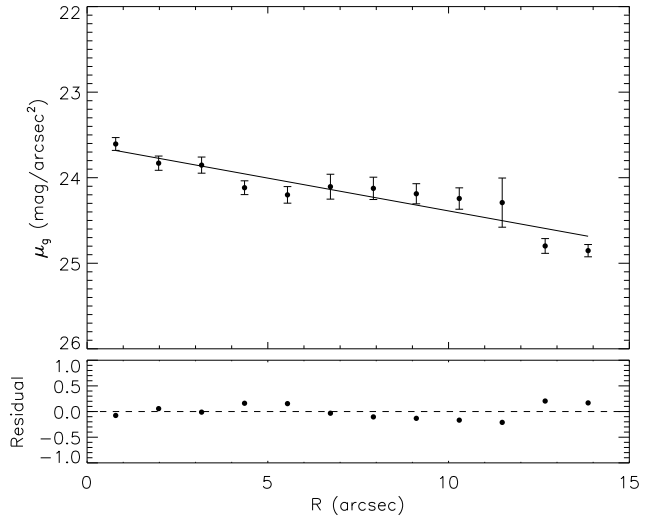}
\caption{The surface brightness profile of KKR 17. The solid line shows the best-fitted single exponential component was shown in solid line. The best-fit is consistent with the outer region of galaxy and GALFIT decomposition results.\label{fig:sb}}
\end{figure}

\subsection{Surface Brightness}
To obtain the surface brightness (SB) profile of KKR 17, we first adopted a geometric center from NED and decomposed the galaxy structure using GALFIT \citep{Peng2002}. Since KKR 17 was bulgeless (S\'{e}rsic index $<$ 1), we fitted the galaxy with the single exponential profile. We obtained the semi-axis ratio of b/a = 0.92, the inclination angle of 34.90 degree and the disk scale length of $r_s$ = 7.52 arcsec. The galactic scale length is equivalent to the linear scale length of 4.20 kpc at this redshift, which was comparable to nearby giant LSBG Malin 1 \citep{Barth2007}. Then, we performed the concentric elliptical aperture photometry according to the GALFIT's results. The surface brightness profile of KKR 17 in the $g$-band was shown in Figure~\ref{fig:sb} and the color of $g-r$ was $\sim$ 0.15 mag. The best-fitted single exponential profile indicated that KKR 17 was a bulgeless galaxy.

In order to obtain the central surface brightness, we converted the SDSS magnitude to Johnson B-band magnitude (Lupton 2005). Then we use the following relation from \citet{Galaz2011} to calculate the surface brightness of KKR 17:

\begin{equation}
\mu_0 (B) = B + 2.5 \log(2{\pi}a^2) + 2.5 \log(b/a) - 10 \log(1+z)
\end{equation}

The values of the semi-axis ratio and redshift are taken from the GALFIT results, and the central surface brightness of KKR 17 was found to be $\mu_0(B)$ = 24.14 $\pm$ 0.03 mag sec$^{-2}$. 

\begin{figure*}
\includegraphics[width=110mm]{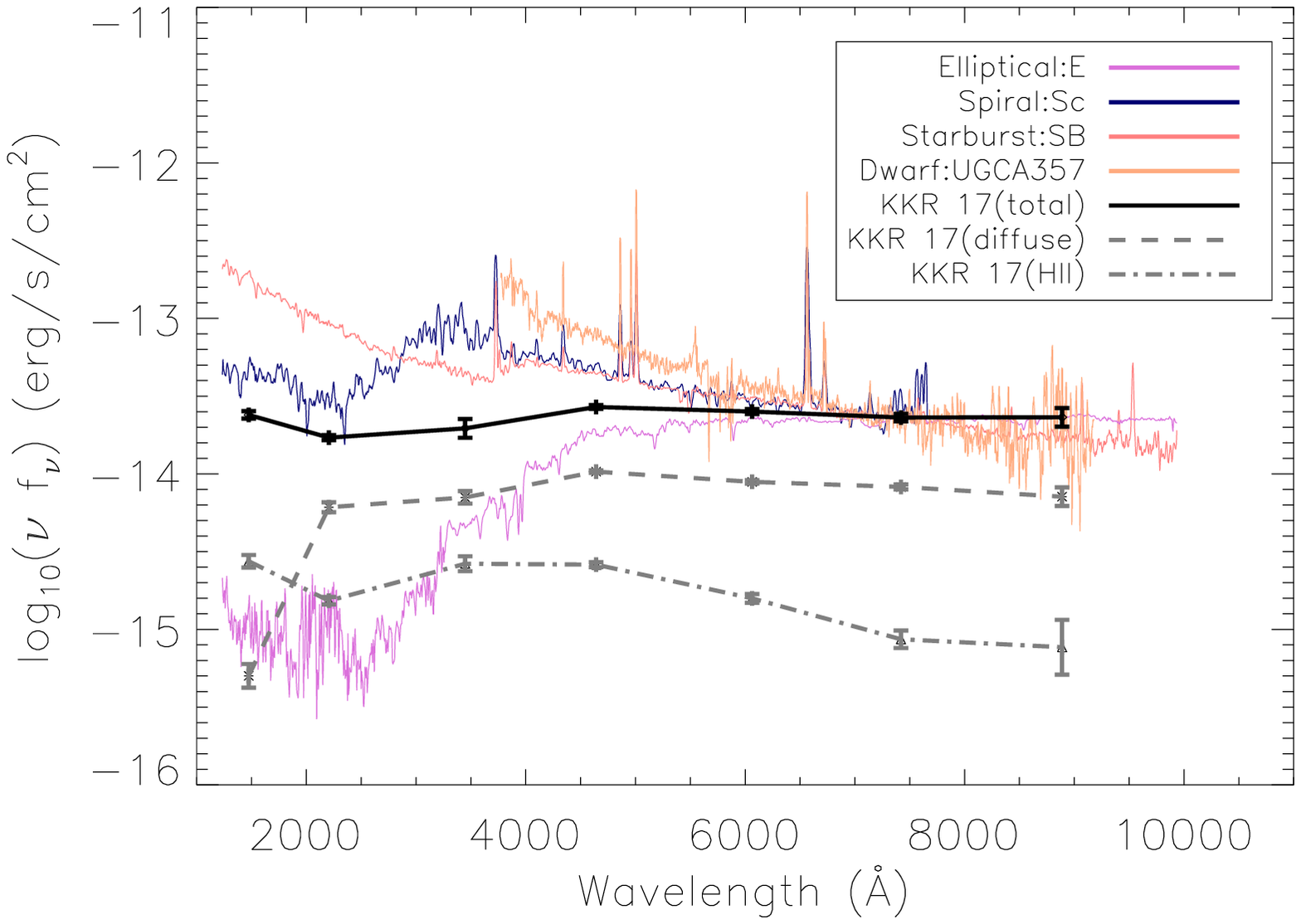}
\caption{The spectral energy distributions of KKR 17, which included the global, diffuse and bright H\Rmnum{2} region. The templates of different type galaxies (E-Sc) are obtained from the Kinney-Calzetti Spectral Atlas of Galaxies \citep{Kinney1996}. The dwarf template is the spectrum of the typical LSBG UGCA357 \citep{vanZee1997}. This figure shows that KKR 17 contains a large fraction of old stellar population, which is comparable to an elliptical galaxy, and the fraction of the young population, however, is lower than UGCA357, but comparable to a Sc-type spiral galaxy.  \label{fig:sed}}
\end{figure*}

\begin{figure*}
\includegraphics[width=110mm]{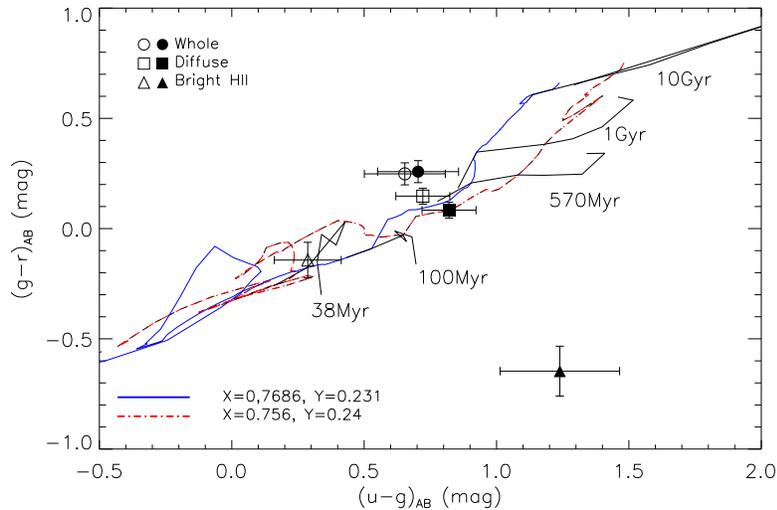}
\caption{The \emph{ugr} color-color diagram of the entire galaxy KKR 17 (\emph{Circle}), H\Rmnum{2} region (\emph{Square}) and diffuse region (\emph{Triangle}) with error bar. \emph{Open symbols}: the regions without emission lines ([O\Rmnum{3}] \& H$\alpha$) subtracted, \emph{solid symbols}: the regions with emission lines subtracted (more description see \SS3.3). The stellar evolution tracks were using Charlot \& Bruzual (CB07) SSP models under the instantaneous SF law. The stellar metallicity were adopted to Z = 0.004, which is the best model for the gas metallicity. \emph{Blue solid line:} the stellar track with X = 0.7686 and Y = 0.231, \emph{red dotted-dash line:} the stellar track with X = 0.756 and Y = 0.24, \emph{Black solid lines:} the isochrone line of 38 Myr, 570 Myr and 10 Gyr, respectively. \label{fig:stellartrack}}
\end{figure*}

\subsection{Stellar Population}
The color of a galaxy could reveal its star formation history (SFH), since the older stellar population would have the redder color, and the younger stellar population would have a bluer color. Previous study found the ages of blue LSBGs span a wide range, from young (2 Gyr) \citep{Zackrisson2005} to old ($\sim$ 7 Gyr) \citep{Jimenez1998}, which depends on the physical conditions both in the disk and SFR \citep{Vorobyov2009}. One of the explanations for these blue LSBGs was that they may be formed by the episodic star formation process (e.g. \citealt{McGaugh1994, Gerritsen1999,Bothun1997}). The episodic star formation process means the stars could form through a discrete process in small perturbation. The star formation process could carry out both active and quiescent periods in the galaxy and last in a few Gyrs. The global stellar population was contributed by the low SFR, and the bright H\Rmnum{2} region was formed by the episodic star formation process. KKR 17 is classified as very blue LSBG, based on the definition of \citet{Oneil1997} (U-B and B-V colors of KKR 17 were 0.25 and 0.27, respectively). It had a significant bright H\Rmnum{2} region but a faint, diffuse region, which would be a perfect testbed for the theoretical models of LSBGs.

The FUV-to-optical SED of the total, diffuse region, and H\Rmnum{2} knot of KKR 17 were shown in Figure~\ref{fig:sed}. We compared the SEDs of KKR 17 with the different morphological templates to reveal the stellar population. We adopted the SED templates of elliptical, late-type spiral galaxies, and low dust extinction starburst galaxies (Reddening E(B-V) $<$ 0.1) from \citet{Kinney1996}, and also took one spectrum of typical low surface brightness galaxy UGCA357 \citep{vanZee1997}. We normalized all fluxes of templates and KKR 17 in the central wavelength of i-band filter, which can avoid the strong emission of [O\Rmnum{3}] and H$\alpha$. The total continuum of KKR 17 seemed to be flatter compared to the ones of late-types and starbursts, and also revealed a hybrid stellar population, which could be the population combination of the spiral Sc galaxies and ellipticals. The H\Rmnum{2} knot was a typical young-star-dominated region. However, the stellar populations of diffuse region was quite different from that of H\Rmnum{2} knot. The diffuse region showed a flux jump between NUV and FUV, which is a typical spectral feature for A-type stars \citep{Gulati1994}. Thus the diffuse region may be dominated by A-type stars. Hence, the diffuse region is the most contributor to the mass of the galaxy, which was consistent with NIR 3.4 \micron~image. 

To derive the visible stellar population in a more precise way, we adopted the stellar evolutionary tracks for single stellar population (SSP) with Salpeter IMF \citep{Salpeter1955}. The stellar evolution tracks were calculated based on Charlot \& Bruzual (CB2007) SSP models under the instantaneous SF law, which were effective to the episodic star formation and instantaneous starburst. For KKR 17, the best metallicity O/H value was adopted to be Z = 0.004 (See \S3.5). Figure~\ref{fig:stellartrack} showed the \emph{ugr} color-color diagram of the entire galaxy, H\Rmnum{2} region, and diffuse region. We considered both the cases with and without emission lines subtracted. The integrated ($u-g$), ($g-r$) colors of entire KKR 17 and diffuse region correspond to ages of $\sim$ 600 Myr.  In addition, the H\Rmnum{2} region is much younger than diffuse region, and the intergrated colors revealed that the age of stellar population of this region was $\sim$40 Myr. However, the age of H\Rmnum{2} region was hard to be defined after emission lines was subtracted because the emission lines in H\Rmnum{2} region were strong. As a add, it was hard to detect the continuum in H\Rmnum{2} region. 

The flux over the wavelength range of H\Rmnum{2} in both SED and stellar synthesis model was similar to the ones of OB stars clusters, which to some extent means the H\Rmnum{2} region was undergoing star formation. However, the existence of large fraction of medium age ($\sim$ 600 Myr) stars in diffuse region may also be seen as the hint of different star formation histories in two regions.  

\subsection{Stellar Mass}
The stellar mass of galaxy is a crucial fundamental parameter for galaxy evolution. The mass-to-light ratio relation and single near-infrared luminosity can be used as useful estimators for the stellar masses. All the methods of constraining the stellar mass of galaxies suffers from a uncertainty of a factor of two because of large uncertainty in the contribution fraction by the thermally pulsing asymptotic branch (TP-AGB) stars in the stellar evolution model. The TP-AGB stars have been found to dominate at some evolutionary phases for galaxies \citep{Bruzual2007}. Many evolutionary tracks are biased when the TP-AGB contribution in stellar population synthesis models is considered. For example, the galaxy masses derived from \citealt{Bruzual2011} (hereafter CB11) and \citealt{Bruzual2007v2} (hereafter CB07) model are much lower than that derived from \citealt{Bruzual2003} (hereafter BC03). In this paper, three different methods were adopted to check the consistency of the stellar mass of KKR 17.

\textbf{1. Optical Mass-to-light Ratio:} The mass-to-light ratio (M/L) can provide an effectively way to estimate stellar masses with small change from model to model. The value of the mass-to-light ratio properly falls within the reasonable range of IMF. \citet{Cole2001} and \citet{Bell2003} derived the M/L, and studied the stellar mass function by combining the near-infrared and optical photometries in normal-galaxies-dominated sample. \citet{deBlok2001} suggested the B-band stellar mass-to-light ratio (M/L$_{\rm{B}}$) was around 1.4 for LSBGs. Under this assumption, the stellar mass was estimated based on the B-band absolute magnitude in \citet{Freeman1970}, and the obtained stellar mass for KKR 17 is 1.32 $\times$ 10$^9$ M$_{\odot}$ by using M/L$_{\rm{B}}$. 

\textbf{2. Optical Colors:} We also adopted the optical ($g-r$) color from \citet{Bell2003} to check the stellar mass derived from M/L$_{\rm{B}}$. We used the following conversion:
\begin{equation} \log  \frac{M_{\star}}{M_{\odot}} = －0.4(M_{r,\rm{AB}}-4.67) + [a_r + b_r(g-r)_{\rm{AB}}+0.15] \end{equation}
where $M_{r,\rm{AB}}$ was the $r$-band absolute magnitude, and ($g-r$)$_{\rm{AB}}$ was the rest-frame color in the AB magnitude system. The coefficients $a_r$ and $b_r$ were adopted -0.306 and 1.097, respectively. The derived stellar mass of KKR 17 with the relation above was 1.22 $\times$ 10$^9$ M$_{\odot}$, which is close to previous method. As we can see, both results showed that the stellar mass of the galaxy was more than 2 times lower than its mass of neutral hydrogen, which indicates that KKR 17 is dominated by neutral hydrogen. The gas fraction and gas ratio of KKR 17 are $f_{gas} \equiv \log (M_{gas}/(M\star+M_{gas})) = -0.08 $, and $\log (M_{gas}/M\star) = 0.71 $, respectively, which is a factor of 2 below the average gas fraction ($\log (M_{gas}/M\star) \sim 1.5$) in 40\% Arecibo ALFALFA total sample \citep{Huang2012}.

\textbf{3. Monochromatic 3.4\micron~luminosity:} The monochromatic 3.4\micron~luminosity is a convenient way to estimate stellar mass since it has a less dependence on star formation history and massive stars. Previous studied have found the 3.4 $\mu$m luminosity could be a stellar mass tracer of galaxies \citep{Wen2013}, and we then used the following conversion method:
\begin{equation} \log \frac{M_{\star}}{M_{\odot}} = (-0.040 \pm 0.001) + (1.120 \pm 0.01) \times \log (\frac{\nu L_{\nu}(3.4~\mu m)}{L_{\odot}}) \end{equation}
The stellar mass of KKR 17 was 3.77 $\times$ 10$^8$ M$_{\odot}$, which only one-third of the stellar mass derived from the M$_{\rm{B}}$/L.

The stellar masses derived from the optical and near-infrared luminosity are quite different. The possible explanation was that the stellar mass estimate depended on TP-AGB model. The near-infrared luminosity in Method 3 may underestimate the stellar mass. In addition, considering Method 2 is less affected by the assumption compared to Method 1, and our adopted stellar mass of KKR 17 in this paper is $1.22\times 10^{9}~M_{\odot}$.

\begin{figure}
\includegraphics[width=0.99\columnwidth]{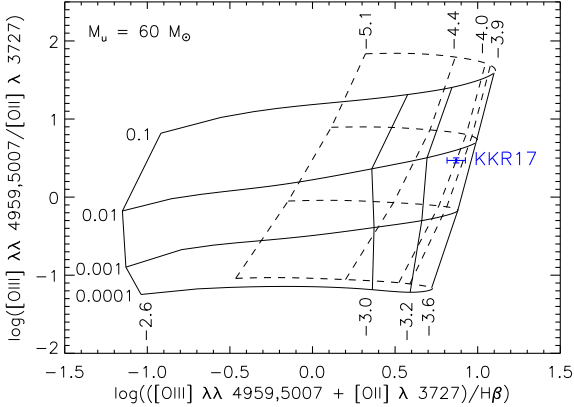}
\caption{Diagnostic diagram of oxygen emission-line ratios compared with the photoionization models of \citet{McGaugh1991}. The position of KKR 17 indicates an oxygen abundance of 12 + $\log$(O/H) = 8.0 $\pm$ 0.1. \label{fig:abundant}}
\end{figure}

\subsection{Oxygen Abundance}
\label{sec:oxygen}
Generally, LSBGs are found to be metal-poor ($Z<1/3Z_{\odot}$) \citep{McGaugh1994b}. Some of them are the very metal-poor extragalactic objects discovered so far. The signal-to-noise ratio (S/N) of [O\Rmnum{3}]$\lambda$4363 of KKR 17 was not high enough, so we only calculated the oxygen abundance for the brightest H\Rmnum{2} region by using the strong-line method described in \citet{McGaugh1991} (hereafter the McGaugh method) as following: \\
\\
$R_{23}$: \begin{math} ([\rm{O}\uppercase\expandafter{\romannumeral3}]\lambda\lambda4959,5007+[\rm{O}\uppercase\expandafter{\romannumeral2}]\lambda3727)/H\beta \end{math} \\
$O_{23}$: \begin{math} ([\rm{O}\uppercase\expandafter{\romannumeral3}]\lambda\lambda4959,5007/[\rm{O}\uppercase\expandafter{\romannumeral2}]\lambda3727) \end{math} \\

Using the $R_{23}$ and $O_{23}$ values, the H\Rmnum{2} region of KKR 17 is overplotted on a grid of theoretical models and the abundance is determined by interpolating between the model points (see Figure~\ref{fig:abundant}). The models used for this work are from \citet{McGaugh1991} in which the massive star produced from the IMF has a upper limit of 60M$_{\odot}$.

The well-studied behavior of the strong oxygen lines (i.e., the $R_{23}$ parameter) with different metallicity has been scaled to the oxygen abundance, using both empirical and theoretical methods (e.g. \citealt{Edmunds1984,McGaugh1991,Kewley2002,Nagao2006}), which is the foundation of all these models. The ionization parameter can also be determined based on the additional parameter $O_{23}$ which leads to a more accurate estimate of the abundance (e.g. \citealt{Kewley2002,Nagao2006}). However, the relationship between $O_{23}$ and $R_{23}$ is not unique: there exist two branches of models, (1) oxygen abundance decreases with an increase in $R_{23}$ as `high-metallicity' branch and (2) oxygen abundance increases with an increase in the $R_{23}$ as `low-metallicity' branch. In brief, each point on the grid of McGaugh's models can lead to two different possible metallicity values.

In order to break the degeneracy, the [N\Rmnum{2}]/[O\Rmnum{2}] ratio is used as a diagnostic to determine whether the H\Rmnum{2} region belongs to the `high-' or `low-' metallicity branch. For KKR 17, the H\Rmnum{2} region is located on the low-metallicity branch since [N\Rmnum{2}]/[O\Rmnum{2}] $<$ -1. and the metallicity is 12+$\log$(O/H) = 8.0 $\pm$ 0.1.

\begin{figure*}
\includegraphics[width=130mm]{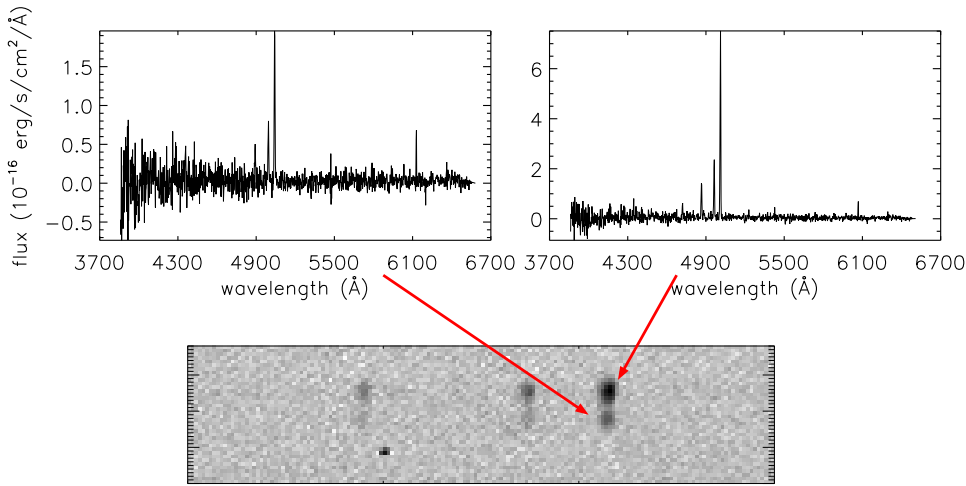}
\caption{The substructure spectra of H\Rmnum{2} region. Left: bottom aperture ; right: upper aperture. The H$\beta$ and [O\Rmnum{3}]$\lambda$$\lambda$ 4959,5007 had slightly different velocities in two regions. Also, the ratios of [O\Rmnum{3}]$\lambda$$\lambda$ 4959,5007 in two regions were slightly different. The flux units were erg/s/cm/\AA~in the spectra. \label{fig:multicomponents}}
\end{figure*}

\subsection{Star Formation Rates }
\label{sec:quantify_sfr}
Star formation rate (SFR) is an important parameter to describe the star formation activities as well as the evolutionary history of galaxies. SFR tracers are explored in a wide wavelength range from UV to sub-millimeter (e.g. \citet{Kennicutt1998,Wu2005,Zhu2008,Calzetti2010}), especially in IR-band since it is closely related to H\Rmnum{2} region. KKR 17 is a typical low surface brightness galaxy, since the IR emissions was contaminated by the nearby bright stars. We adopted the narrow-band H$\alpha$ as the star formation tracer in this paper.

The total H$\alpha$ luminosity of KKR 17 was determined from the narrow-band image with both background and continuum subtracted, by integrating the flux over the aperture that encloses the entire galaxy. The total flux of H$\alpha$ we have obtained in KKR 17 was (1.51$\pm$0.26) $\times$ 10$^{-14}$ erg/s/cm$^2$, or L(H$\alpha$) = (2.65$\pm$0.46) $\times$ 10$^{40}$ erg/s for a distance of 120.8 $\pm$ 8.5 Mpc. In order to estimate the global SFR, we adopted the following relation~\citet{Kennicutt1998}:\\

\begin{equation} \rm{SFR}~(\rm{M}_{\odot}/\rm{yr}) = 7.9 \times 10^{-42}~L(H\alpha) (\rm{ergs/s}) \end{equation} 

Then, it yields a total SFR of 0.21$\pm$0.04 M$_{\odot}$/yr, and the specific star formation rate (sSFR) is 1.80 $\times$ 10$^{-10}$yr$^{-1}$, using the \citet{Bell2003}'s result, which means the time scale to form the total mass of KKR 17 under such SFR is 10$^{10}$ yr. The sSFR value is slightly higher to the one for the typical E/S0 galaxies inferred for IR-selected sample~\citet{Lam2013}. Therefore, we conclude that the star formation process of KKR 17 is still in the episodic star formation process.

The flux coming from the H\Rmnum{2} region takes roughly 43\% of total H$\alpha$ flux in the H$\alpha$ narrow-band image, which was consistent with the results of \citet{Schombert2013}. The SFR in this H\Rmnum{2} region is 0.09$\pm$0.02 M$_{\odot}$/yr, which is similar to the SFRs of young stellar clusters in M51 \citep{Calzetti2005}, a major merging system Arp 24 \citep{Cao2007},and a minor merger galaxy NGC7479 \citep{Zhou2011}.

\section{Discussion}
\subsection{Bright H\Rmnum{2} region}
KKR 17 showed a clearly bright H\Rmnum{2} region, on its east side, dominated by the massive O stars. The H\Rmnum{2} region of KKR 17 was more compact and luminous than the \citet{Schombert2013}'s sample. The total H$\alpha$ luminosity of this H\Rmnum{2} region was approximately to $\log$ L(H\Rmnum{2}) = 40.06 erg/s, within the radius of  $\sim$ 1 kpc. If the luminosity of a single O7V star in a H\Rmnum{2} region is $\log$ L(H\Rmnum{2}) = 37.0 \citep{Schombert2013, Werk2008}, then thousands of O stars may exist in the H\Rmnum{2} region, and they contributed $\sim$ 50\% to H$\alpha$ luminosity in KKR 17, which is consistent with the previous results~\citet{vanZee2000} and \citet{Schombert2013}.

The emission lines of H\Rmnum{2} region in KKR 17 showed the star formation processes may be complicated. Figure~\ref{fig:multicomponents} showed two components in the same region of KKR 17. The physical separation of the two components was $\sim$1.8 kpc, which similar to the size of the H\Rmnum{2} region. The components showed a small velocity difference of $\sim$60 km/s identified by [O\Rmnum{3}] emission lines, and $\sim$90 km/s by H$\beta$ emission line. Also, the metallicities were slightly different in terms of $R_3$ index, which can be easily obtained as a rough metallicity estimator but with relative large uncertainty. The parameter $R_3$ can be obtained through the simple relation of $R_3$ = 1.35 $\times$ I$_{[\rm{O}\Rmnum{3}]5007}$ \citep{Vacca1992}, where $I_{[\rm{O}\Rmnum{3}]5007}$ is the flux of [O\Rmnum{3}]. Once the value of $R_3$ is known, the metallicity of 12 + $\log$ (O/H) can be derived with the equation (7) in \citet{Vacca1992}. For the two flux values of [O\Rmnum{3}], which correspond to different components of H\Rmnum{2} region, the values of the metallicity are $\sim$8.2 and $\sim$8.3, respectively, and the difference is 0.1 dex. In contrast, it is noted that the systematic uncertainty is larger, 0.5 dex.

The small differences in the velocity and metallicity of H\Rmnum{2} components can be attributed to the internal episodic star formation activities. In fact, the young stellar clusters (YSCs), often found in HSBGs, can generate shocks, which can be used to explain those differences \citep{Elmegreen2004}. \citet{Schombert2013} pointed out that the star formation processes in LSBGs are the same as the one in HSBGs. In addition, the recent research on YSCs in HSBGs found that YSCs are readily formed in a cluster complex than in an isolated environment \citep{Bastian2005}. The strong stellar wind from clustering massive stars could lead to those differences in YSCs. The shock-accelerated superwind was detected by nearby galaxies M82 and NGC839 \citep{Rich2010}. However, this was not supported by our spectra, we did not detect any line as the strong shock tracer, i.e., the [SII] lines was weak and the [N\Rmnum{2}] lines was an upper limit detection. 

Another explanation is that each component had different origins. It was possible that the bright H\Rmnum{2} region was a dwarf galaxy fell into KKR 17 from a minor merger. The blue color of H\Rmnum{2} region, $g-i = -0.2$, is similar to the one for a irregular galaxy \citep{Fukugita1995}, and its absolute magnitude (M$_B$ $\sim$ -15) resided at the faint end of luminosity function for the local extreme low-luminosity galaxies.  Moreover, the unusual colors of B-V = 0.27 and U-B = 0.21 might also imply the existence of an infalling dwarf galaxy in KKR 17. The firm conclusion may still need a higher spatial resolution study for YSCs in KKR 17, e.g., the future IFU spectroscopy.

\subsection{The Possible Evolution Scenarios of KKR 17}
The LSBGs are usually thought to be evolving slowly across the Hubble time compared with their high surface brightness counterparts. The slow evolution may be connected with the small gas surface density\citep{vanderHulst1993}. If a LSBGs is lack of neighbors over small and intermediate space scales, the absence of gravitational interaction will be able to keep the gas in a stable situation\citep{Bothun1993}. Also the other possibilities for the slow evolution of LSBGs could be due to the large dark matter halo, or low metallicity and dust content, or different IMF\citep{Mihos1997, Oneil1997}. 

The blue, gas-rich irregular galaxies, KKR 17, is located in Local Void in Hercules-Aquila direction. Theoretically, it may be far away from interaction in a long time scale, which leads to a low star formation rate. The previous studies on the SFH of IZw 18 showed that the continuously long-time low star formation activity can build up the stellar mass of galaxy \citep{Aloisi1999,Legrand2000,Annibali2013}. Although the slow star formation process can not be confirmed, we found that the KKR 17 experienced a short-duration star formation roughly 600 Myr ago. \citet{Kim2007} pointed out the LSBGs were fast evolving in recent $\sim$1 Gyr. Their results are in favor of the episodic star formation scenario \citep{McGaugh1992}. The medium-age stellar population of diffuse region in KKR 17 may be dominant by A-type stars, which is consistent with the \citet{Kim2007}'s result.

Except the star formation process occurred 600 Myr ago, the bright H\Rmnum{2} region is currently undergoing the strong star formation in KKR 17. The strong emission line features of H\Rmnum{2} region indicated the existence of a large fraction of O stars. The luminosity of H\Rmnum{2} region (L(H$\alpha$) $>$ 10$^{40}$ erg/s) was unusually high to a LSBG \citep{Helmboldt2009, Schombert2013}.  As suggested by \citet{Bastian2006}, the star formation in such complicated YSC could be triggered by an external perturbation. If the perturbation was from an nearby galaxy, we should find the evidence in observations. However, we don't found any objects from our FUV-to-NIR images. It is an isolated galaxy without any obvious large galaxies within a angular radius of 8\arcmin, which is equivalent to $\sim$250 kpc at its redshift. Based on the discussion above, the multiple components of H$\beta$ and [O\Rmnum{3}] emission lines may be a clue for the minor merger remnant, or the H\Rmnum{2} region was a dwarf galaxy, as discussed in Section \S4.1, and experienced a minor merger. The deep interferometric imaging to KKR 17 with high spatial resolution and velocity dispersion may clean up the confusion, revealing its real star formation activity. 

\section{Summary}
\label{sec:discussion}
In this paper, we present the results on metallicity and SFH of low surface brightness galaxy, KKR 17, with ground-based optical images and spectra combined with space telescope archival data, and they are summarized as the following:

1. KKR 17 is a HI-dominated low surface brightness galaxy of which the M(HI) = 4.37 $\times$ 10$^9$ M$_{\odot}$, and the stellar mass is only about several 10$^8$ M$_{\odot}$ to 10$^9$ M$_{\odot}$ with a central surface brightness of $\mu_0(B)$ = 24.14 $\pm$0.03 magsec$^{-2}$.

2. The metallicity of the entire KKR 17 is 12 + $\log$(O/H) = 8.0 $\pm$ 0.1 with \citet{McGaugh1991}'s model. However, the multiple components have a slightly different metallicities.

3. The global SFR of KKR 17 is 0.21$\pm$0.04 M$_{\odot}$/yr, which is $\sim$1/5 of our Milky Way's. The sSFR is 1.80 $\times$ 10$^{-10}$ yr$^{-1}$ that leads to 10$^{10}$ yr to form the total stellar mass of KKR 17. This sSFR was similar to the one for E/S0 galaxies. Therefore, KKR 17 is in quiescent star formation stage.

4. The fits to the optical SED and color-color diagrams of the diffuse and H\Rmnum{2} regions have revealed different stellar populations, which may represent the distinctive history of the star formation activities.  

\section*{Acknowledgements}
We would like to thank the staff of the 2.16m telescope at Xinglong Observatory for their excellent support during our observing runs. We would also like to thank Dr. L. J. Gou for his kindly help thoughout the paper. We would like to acknowledge the anonymous referee for his/her helpful suggestions and comments.

This project is supported by the National Natural Science Foundation of China (Grant No.11173030), the China Ministry of Science and Technology under the State Key Development Program for Basic Research (2012CB821800, 2014CB845705), the National Natural Science Foundation of China (Grant Nos. 11225316, 11078017, 11303038, 10833006, 10978014, 10773014 and 11403061), the Key Laboratory of Optical Astronomy, the National Astronomical Observatories, Chinese Academy of Sciences. Supported by the Strategic Priority Research Program `The Emergence of Cosmological Structures' of the Chinese Academy of Sciences, Grant No. XDB09000000.

We thank the work of the entire ALFALFA collaboration team in observing, flagging, and extracting the catalog of galaxies used in this work. We acknowledge NASA's support for construction, operation, and science analysis for the \emph{GALEX} mission.

Funding for the creation and distribution of the SDSS archive has been provided by the Alfred P. Sloan Foundation, the Participating Institutions, the National Aeronautics and Space Administration, the National Science Foundation, the U.S. Department of Energy, the Japanese Monbukagakusho, and the Max Plank Society. The SDSS website is http://www.sdss.org. The SDSS is managed by the Astrophysical Research Consortium (ARC) for the Participating Institutions.

This publication makes use of data products from the Wide-field Infrared Survey Explorer, which is a joint project of the University of California, Los Angeles, and the Jet Propulsion Laboratory/California Institute of Technology, funded the National Aeronautics and Space Administration.


\bsp

\label{lastpage}


\begin{thebibliography}{ }

\bibitem[\protect\citeauthoryear{Aloisi et al.}{1999}]{Aloisi1999}
Aloisi, A., Tosi, M., Greggio, L., 1999, ApJ, 118, 302

\bibitem[\protect\citeauthoryear{Annibali et al.}{2013}]{Annibali2013}
Annibali, F., et al., 2013, ApJ, 146, 144

\bibitem[\protect\citeauthoryear{Barth et al.}{2007}]{Barth2007}
Barth A. J., 2007, AJ, 133, 1085

\bibitem[\protect\citeauthoryear{Bastian et al.}{2005}]{Bastian2005}
Bastian, N., Gieles, M., Efremov, Yu. N., Lamers, H. J. G. L. M., 2005, A\&A, 443, 79

\bibitem[\protect\citeauthoryear{Bastian et al.}{2006}]{Bastian2006}
Bastian, N., Emsellem, E., Kissler-Patig, M., Maraston, C., 2006, A\&A, 445, 471

\bibitem[\protect\citeauthoryear{Bergmann et al.}{2003}]{Bergmann2003}
Bergmann, M. P., J{\o}rgensen, I., Hill, G. J., 2003, AJ, 125, 116

\bibitem[\protect\citeauthoryear{Bell et al.}{2003}]{Bell2003}
Bell, E. F., McIntosh, D. H., Katz, N., Weinberg, M. D., 2003, ApJS, 149, 289

\bibitem[\protect\citeauthoryear{Bothun et al.}{1992}]{Bothun1992}
Bothun, G. D., et al., 1992, ApJ, 395, 347

\bibitem[\protect\citeauthoryear{Bothun et al.}{1993}]{Bothun1993}
Bothun, G. D., Schombert, J. M., Impey, C. D., Sprayberry, D., McGaugh, S. S., 1993, AJ, 106, 530

\bibitem[\protect\citeauthoryear{Bothun, Impey \& McGaugh}{1997}]{Bothun1997}
Bothun, G., Impey, C., McGaugh, S., 1997, PASP, 109, 745

\bibitem[\protect\citeauthoryear{Bruzual \& Charlot}{2003}]{Bruzual2003}
Bruzual, G., Charlot, S. 2003, MNRAS, 344, 1000

\bibitem[\protect\citeauthoryear{Bruzual}{2007A}]{Bruzual2007}
Bruzual, G., 2007, ASPC, 374, 303

\bibitem[\protect\citeauthoryear{Bruzual}{2007B}]{Bruzual2007v2}
Bruzual, G. 2007, IAUS, 241, 125

\bibitem[\protect\citeauthoryear{Bruzual}{2011}]{Bruzual2011}
Bruzual, G. 2011, RMxAC, 40, 36

\bibitem[\protect\citeauthoryear{Caldwell \& Bothun}{1987}]{Caldwell1987}
Caldwell, N., Bothun, G. D., 1987, AJ, 94, 1126

\bibitem[\protect\citeauthoryear{Calzetti et al.}{2005}]{Calzetti2005}
Calzetti, D., et. al., 2005, ApJ, 633, 871

\bibitem[\protect\citeauthoryear{Calzetti et al.}{2010}]{Calzetti2010}
Calzetti, D., et al., 2010, ApJ, 714, 1256

\bibitem[\protect\citeauthoryear{Cao \& Wu}{2007}]{Cao2007}
Cao, C., \& Wu, H., 2007, AJ, 133, 1710


\bibitem[\protect\citeauthoryear{Cole et al.}{2001}]{Cole2001}
Cole, S. et al., 2001, MNRAS, 326, 255


\bibitem[\protect\citeauthoryear{de Blok, McGaugh \& Rubin}{2001}]{deBlok2001}
de Blok, W. J. G., McGaugh, S. S., \& Rubin, V. C., AJ, 2001, 122, 2396

\bibitem[\protect\citeauthoryear{Disney}{1976}]{Disney1976}
Disney, M.J., 1976, Nature, 263, 573


\bibitem[\protect\citeauthoryear{Edmunds \& Pagel}{1984}]{Edmunds1984}
Edmunds, M. G., \& Pagel, B. E., 1984, MNRAS, 211, 507

\bibitem[\protect\citeauthoryear{Elmegreen}{2004}]{Elmegreen2004}
Elmegreen, B.G., 2004, APSC, 322, 277

\bibitem[\protect\citeauthoryear{Freeman}{1970}]{Freeman1970}
Freeman, K. C., 1970, ApJ, 160, 811

\bibitem[\protect\citeauthoryear{Fukugita}{1995}]{Fukugita1995}
Fukugita, M., Shimasaku, K., Ichikawa, T., 1995, PASP, 107, 945

\bibitem[\protect\citeauthoryear{Galaz et al.}{2011}]{Galaz2011}
Galaz, G., Herrera-Camus, R., Garcia-Lambas, D., Padilla, N., 2011, ApJ, 728, 74

\bibitem[\protect\citeauthoryear{Gerritsen \& de Blok}{1999}]{Gerritsen1999}
Gerritsen, J. P. E., de Blok, W. J. G., 1999, A\&A, 342, 655

\bibitem[\protect\citeauthoryear{Giovanelli et al.}{2005a}]{Giovanelli2005a}
Giovanelli, R., et al. 2005, AJ, 130, 2598

\bibitem[\protect\citeauthoryear{Gulati et al.}{1994}]{Gulati1994}
Gulati, R. K., Gupta, R., Gothoskar, P., Khobragade, S., VA, 38, 293

\bibitem[\protect\citeauthoryear{Haynes et al.}{2011}]{Haynes2011}
Haynes, M. P., et al., 2011, AJ, 142, 170

\bibitem[\protect\citeauthoryear{Huchtmeier, Karachentsev \& Karachentseva}{2000}]{Huchtmeier2000}
Huchtmeier, W. K., Karachentsev, I. D., Karachentseva, V. E., 2000, A\&AS, 147, 187

\bibitem[\protect\citeauthoryear{Huang et al.}{2012}]{Huang2012}
Huang, S. Haynes, M. P., Giovanelli, R., Brinchmann, J., 2012, ApJ, 756, 113

\bibitem[\protect\citeauthoryear{Helmboldt et al.}{2009}]{Helmboldt2009}
Helmboldt, J. F., Walterbos, R. A. M., Bothun, G. D., O'Neil, K., Oey, M. S., 2009, MNRAS, 393, 478

\bibitem[\protect\citeauthoryear{Impey et al.}{1996}]{Impey1996}
Impey, C. D., Sprayberry, D., Irwin, M. J., Bothun, G. D., 1996, ApJS, 105, 209

\bibitem[\protect\citeauthoryear{Impey \& Bothun}{1997}]{Impey1997}
Impey, C., Bothun, G., 1997, ARAA, 35, 267

\bibitem[\protect\citeauthoryear{Jimenez et al.}{1998}]{Jimenez1998}
Jimenez, R., Padoan, P., Matteucci, F., Heavens, A. F. 1998, MNRAS, 299, 123

\bibitem[\protect\citeauthoryear{Karachentseva, Karachentsev \& Richter}{1999}]{KKR1999}
Karachentseva, V. E., Karachentsev, I. D., \& Richter, G. M., 1999, A\&AS, 135, 221

\bibitem[\protect\citeauthoryear{Kennicutt}{1998}]{Kennicutt1998}
Kennicutt, R. C. Jr., 1998, ARAA, 36, 189

\bibitem[\protect\citeauthoryear{Kewley \& Dopita}{2002}]{Kewley2002}
Kewley, L. J., Dopita, M. A., 2002, ApJS, 142, 35

\bibitem[\protect\citeauthoryear{Kim}{2007}]{Kim2007}
Kim, J. H., 2007, Ph.D Thesis

\bibitem[\protect\citeauthoryear{Kinney et al.}{1996}]{Kinney1996}
Kinney, A. L., et al., 1996, ApJ, 467, 38

\bibitem[\protect\citeauthoryear{Lam et al.}{2013}]{Lam2013}
Lam, M. I., Wu, H., Zhu, Y.-N., Zhou, Z.-M., 2013, RAA, 13, 179

\bibitem[\protect\citeauthoryear{Legrand et al.}{2000}]{Legrand2000}
Legrand, F., Kunth, D., Roy., J.-R., Mas-Hesse, J. M., \& Walsh, J. R., 2000, MNRAS, 355, 891

\bibitem[\protect\citeauthoryear{Makarov, Karachentsev \& Burenkov}{2003}]{Makarov2003}
Makarov, D. I., Karachentsev, I. D., Burenkov, A. N., 2003, A\&A, 405, 951

\bibitem[\protect\citeauthoryear{Martin et al.}{2005}]{Martin2005}
Martin, D.C., et al., 2005, ApJ, 619, 1

\bibitem[\protect\citeauthoryear{McGaugh}{1991}]{McGaugh1991}
McGaugh, S. S., 1991, ApJ, 380, 140

\bibitem[\protect\citeauthoryear{McGaugh}{1992}]{McGaugh1992}
McGaugh, S. S., 1992, Ph.D Thesis

\bibitem[\protect\citeauthoryear{McGaugh}{1994}]{McGaugh1994}
McGaugh, S. S., 1994, Nature, 367, 538

\bibitem[\protect\citeauthoryear{McGaugh}{1994}]{McGaugh1994b}
McGaugh, S. S., 1994, ApJ, 426, 135

\bibitem[\protect\citeauthoryear{McGaugh, Schombert, \& Bothun}{1995}]{McGaugh1995}
McGaugh, S. S., Schombert, J. M., Bothun, G. D. 1995, AJ, 109, 2019


\bibitem[\protect\citeauthoryear{Mihos et al.}{1997}]{Mihos1997}
Mihos, J. C., McGaugh S. S., de Blok, W. J. G., 1997, ApJ, 481, 741

\bibitem[\protect\citeauthoryear{Morrissey et al.}{2007}]{Morrissey2007}
Morrissey, P., et al., 2007, ApJS, 173, 682

\bibitem[\protect\citeauthoryear{Nagao et al.}{2006}]{Nagao2006}
Nagao, T., Maiolino, R., \& Marconi, A., 2006, A\&A, 459, 85

\bibitem[\protect\citeauthoryear{O'Neil et al.}{1997}]{Oneil1997}
O'Neil, K., Bothun, G. D., Schombert, J., Cornell, M. E., Impey, C. D., 1997, AJ, 114, 2448



\bibitem[\protect\citeauthoryear{Peng et al.}{2002}]{Peng2002}
Peng, C. Y., Ho, L. C., Impey, C. D.; Rix, H., 2002, AJ, 124, 266

\bibitem[\protect\citeauthoryear{Rich et al.}{2010}]{Rich2010}
Rich, J. A., Dopita, M. A., Kewley, L. J., Rupke, D. S. N., 2010, ApJ, 721, 505

\bibitem[\protect\citeauthoryear{Salpeter}{1955}]{Salpeter1955}
Salpeter, E. E., 1955, ApJ, 121, 161

\bibitem[\protect\citeauthoryear{Schombert \& Bothun}{1988}]{Schombert1988}
Schombert, J. M., Bothun, G. D., 1988, AJ, 95, 1389

\bibitem[\protect\citeauthoryear{Schombert et al.}{1992}]{Schombert1992}
Schombert, J. M., et al., 1992, AJ, 103, 1107

\bibitem[\protect\citeauthoryear{Schombert, McGaugh \& Eder}{2001}]{Schombert2001}
Schombert, J. M., McGaugh, S. S., Eder., J., 2001, AJ, 121, 2420

\bibitem[\protect\citeauthoryear{Schombert, Maciel \& McGaugh}{2011}]{Schombert2011}
Schombert, J., Maciel, T., McGaugh, S., 2011, AdAst, 2011, 12

\bibitem[\protect\citeauthoryear{Schombert, McGaugh \& Maciel}{2013}]{Schombert2013}
Schombert, J. M., McGaugh, S. S., Maciel, T., 2013, AJ, 146, 41

\bibitem[\protect\citeauthoryear{Schombert \& McGaugh}{2014}]{Schombert2014}
Schombert, J. M., \& McGaugh, S. S., 2014, PASA, 31, 11

\bibitem[\protect\citeauthoryear{Skillman, Kennicutt \& Hodge}{1989a}]{Skillman1989a}
Skillman, E. D., Kennicutt, R. C., Hodge, P. W., 1989, ApJ, 347, 875

\bibitem[\protect\citeauthoryear{Skillman, Terlevich \& Melnick}{1989b}]{Skillman1989b}
Skillman, E. D., Terlevich, R., Melnick, J., 1989, MNRAS, 240, 563

\bibitem[\protect\citeauthoryear{Stoughton et al.}{2000}]{Stoughton2002}
Stoughton, C., et al., 2002, AJ, 123, 485

\bibitem[\protect\citeauthoryear{Vacca \& Conti}{1992}]{Vacca1992}
Vacca, W. D., \& Conti, P. S., 1992, ApJ, 401, 543

\bibitem[\protect\citeauthoryear{van der Hulst et al.}{1993}]{vanderHulst1993}
van der Hulst, J. M., Skillman, E. D., Smith, T. R., Bothun, G. D., McGaugh, S. S., de Blok, W. J. G., 1993, AJ, 106, 548

\bibitem[\protect\citeauthoryear{van Zee, Haynes \& Salzer}{1997}]{vanZee1997}
van Zee, L., Haynes, M. P., Salzer, J. J., 1997, AJ, 114, 2497

\bibitem[\protect\citeauthoryear{van Zee}{2000}]{vanZee2000}
van Zee, L., 2000, ApJ, 119, 2757

\bibitem[\protect\citeauthoryear{Vorobyov et al.}{2009}]{Vorobyov2009}
Vorobyov, E. I., Shchekinov, Yu., Bizyaev, D.; Bomans, D., Dettmar, R.-J., 2009, A\&A, 505, 483

\bibitem[\protect\citeauthoryear{Werk et al.}{2008}]{Werk2008}
Werk, J. K., et al., 2008, ApJ, 678, 888

\bibitem[\protect\citeauthoryear{Wen et al.}{2013}]{Wen2013}
Wen, X.-Q., et al., 2013, MNRAS, 433, 2946

\bibitem[\protect\citeauthoryear{Wright et al.}{2010}]{Wright2010}
Wright E. L., et al., 2010, AJ, 140, 1868

\bibitem[\protect\citeauthoryear{Wu et al.}{2005}]{Wu2005}
Wu,H. et al., ApJ, 632, 79

\bibitem[\protect\citeauthoryear{York et al.}{2000}]{York2000}
York, D. G., et al., 2000, AJ, 120, 1579

\bibitem[\protect\citeauthoryear{Zackrisson et al.}{2005}]{Zackrisson2005}
Zackrisson, E.; Bergvall, N.; \"{O}stlin, G., 2005, A\&A, 435, 29

\bibitem[\protect\citeauthoryear{Zhou et al.}{2011}]{Zhou2011}
Zhou, Z-M, Cao, C., Meng, X.-M., Wu, H., 2011, AJ, 142, 38

\bibitem[\protect\citeauthoryear{Zhu et al.}{2008}]{Zhu2008}
Zhu, Y.-N., Wu, H., Cao, C., Li, H.-N., 2008, ApJ, 686, 155

\end{thebibliography}
\end{document}